\documentclass[conference]{IEEEtran}
\usepackage{amsmath,amssymb,amsfonts}
\usepackage{graphicx}
\usepackage{caption}
\usepackage{subcaption}
\usepackage{booktabs}
\usepackage{algorithm}
\usepackage{algpseudocode}
\usepackage{hyperref}
\usepackage{xcolor}
\usepackage{lipsum} 
\usepackage{multirow}
\usepackage{siunitx} 
\usepackage{float}
\usepackage{times}
\usepackage{xcolor}
\definecolor{ABBRed}{RGB}{220,36,31}
\usepackage{tikz}
\usetikzlibrary{shapes.geometric, arrows.meta, positioning, fit, shadows}
\begin{document}

\title{An Explainable Equity-Aware P2P Energy Trading Framework for Socio-Economically Diverse Microgrid}

\author{
    \IEEEauthorblockN{Abhijan Theja, Mayukha Pal\IEEEauthorrefmark{1}}\\
}

\maketitle

\begin{abstract}
Fair and dynamic energy allocation in community microgrids remains a critical challenge, particularly when serving socio-economically diverse participants. Static optimization and cost-sharing methods often fail to adapt to evolving inequities, leading to participant dissatisfaction and unsustainable cooperation. This paper proposes a novel framework that integrates multi-objective mixed-integer linear programming (MILP), cooperative game theory, and a dynamic equity-adjustment mechanism driven by reinforcement learning (RL). At its core, the framework utilizes a bi-level optimization model grounded in Equity-regarding Welfare Maximization (EqWM) principles, which incorporate Rawlsian fairness to prioritize the welfare of the least advantaged participants. We introduce a Proximal Policy Optimization (PPO) agent that dynamically adjusts socio-economic weights in the optimization objective based on observed inequities in cost and renewable energy access. This RL-powered feedback loop enables the system to learn and adapt, continuously striving for a more equitable state. To ensure transparency, Explainable AI (XAI) is used to interpret the benefit allocations derived from a weighted Shapley value. Validated across six realistic scenarios, the framework demonstrates peak demand reductions of up to 72.6\%,  and significant cooperative gains. The adaptive RL mechanism further reduces the Gini coefficient over time, showcasing a pathway to truly sustainable and fair energy communities.
\end{abstract}


\section{Introduction}

The global paradigm shift towards sustainable energy systems has catalyzed the proliferation of community microgrids. These decentralized energy ecosystems enhance grid resilience, optimize local resources, and facilitate the deep integration of Distributed Energy Resources (DERs) like solar photovoltaics (PV) and Battery Energy Storage Systems (BESS) \cite{wang2022transactive}. By enabling participants to act as "prosumers"—both producing and consuming energy—these microgrids foster a collaborative Peer-to-Peer (P2P) energy trading environment, fundamentally altering the traditional, passive role of energy consumers \cite{dwivedi2024evaluation, billah2023decentralized}.

\begingroup\footnotesize
\thanks{(*Corresponding author: Mayukha Pal)}

\thanks{Mr. Abhijan Theja is a Data Science Research Intern at ABB Ability Innovation Center, Hyderabad 500084, India, and also an undergraduate at the Department of Computer Science and Engineering, Indian Institute of Technology Jodhpur, Jodhpur 342037, IN.}

\thanks{Dr. Mayukha Pal is with ABB Ability Innovation Center, Hyderabad 500084, IN, working as Global R\&D Leader – Cloud \& Advanced Analytics (e-mail: mayukha.pal@in.abb.com).}
\endgroup
\vspace{0.5em}

However, this transition introduces formidable operational and social challenges. The inherent intermittency of renewable generation, coupled with heterogeneous consumer load profiles, renders traditional optimization methods inadequate \cite{hannan2020optimized}. While sophisticated techniques like Model Predictive Control \cite{hu2021model} and multi-objective optimization \cite{salehi2022comprehensive} have improved technical efficiency, they frequently neglect the crucial dimension of equity. Similarly, cooperative game-theoretic approaches, such as Shapley value-based cost allocation \cite{churkin2021review}, offer mathematically fair distribution mechanisms but are often decoupled from real-time operational optimization, limiting their practical impact \cite{alam2019networked}.

A critical, yet often overlooked, aspect is the socio-economic diversity within a community. Income and resource disparities play a major role in shaping energy access and affordability, making equity not just a preference but a fundamental aspect of sustainable microgrid design \cite{faisal2018review, sinha2020power}. This concern echoes the Pigou-Dalton principle from economics, which posits that transferring resources from the affluent to the less fortunate enhances overall social welfare \cite{atkinson1970measurement}. This principle finds its modern expression in Social Welfare Functions (SWFs), particularly the Rawlsian SWF, which seeks to maximize the welfare of the least well-off individual—a concept now central to the discourse on energy justice \cite{sen1970collective, rawls1971theory}.

Our work extends the Equity-regarding Welfare Maximization (EqWM) frameworks \cite{li2023decentralized, victor2025integrated} by incorporating Rawlsian fairness directly into a single-level optimization model. The formulation simultaneously addresses system-wide welfare and individual household utilities under equitable pricing contracts.

The primary innovation of this paper lies in addressing the static nature of existing models. Equity is not a one-time calculation but a dynamic state that can erode over time. To this end, we introduce a novel Reinforcement Learning (RL)-driven dynamic equity adjustment mechanism. We formulate the problem as a Markov Decision Process (MDP) where a Proximal Policy Optimization (PPO) agent observes the outcomes of the MILP optimization—specifically, the distribution of costs and renewable energy benefits—and learns to adjust the socio-economic weights applied to each household in the objective function. This adaptive feedback loop allows the system to:
\begin{enumerate}
    \item Learn from Inequality: The RL agent is rewarded for reducing inequality and improving the utility of underprivileged participants.
    \item Adapt Dynamically: Unlike fixed weights, the RL-adjusted weights evolve over time, making the system responsive to persistent disparities.
    \item Promote Long-Term Fairness: The mechanism works to maintain fairness over extended operational periods, fostering sustainable cooperation among diverse participants.
    \item To maintain transparency, we further employed Explainable AI (XAI) to interpret the final benefit allocations. Our framework is rigorously validated across six operational scenarios, demonstrating superior performance in both technical efficiency and equity metrics.
\end{enumerate}

The remainder of this paper is organized to systematically build and validate our proposed framework. Section II details the core methodology, beginning with the system architecture and then presenting the comprehensive MILP formulation, including all operational, economic, and equity-based constraints. Section III provides a deep dive into the dynamic equity adjustment mechanism, explaining the rationale for RL and detailing the MDP formulation and the PPO algorithm. Section IV describes the game-theoretic approach to fair benefit allocation using the weighted Shapley value and presents the integrated algorithm that combines all framework components. Section V presents the simulation setup and provides a thorough analysis of the results across six diverse scenarios, highlighting technical performance and equitable outcomes. Finally, Section VI concludes the paper, summarizing our findings and outlining promising directions for future research.
\section{Methodology and System Model}

This section details the proposed framework for achieving equitable and efficient P2P energy trading in a socio-economically diverse participant microgrid. The methodology integrates three core pillars: (1) a multi-objective Mixed-Integer Linear Programming (MILP) model for operational optimization, (2) a dynamic equity-adjustment layer using Reinforcement Learning (RL), and (3) a fair benefit allocation mechanism based on cooperative game theory and XAI.

\subsection{System Architecture  and Data Flow}
\begin{figure}[H]
    \centering
    \includegraphics[width=0.9\columnwidth]{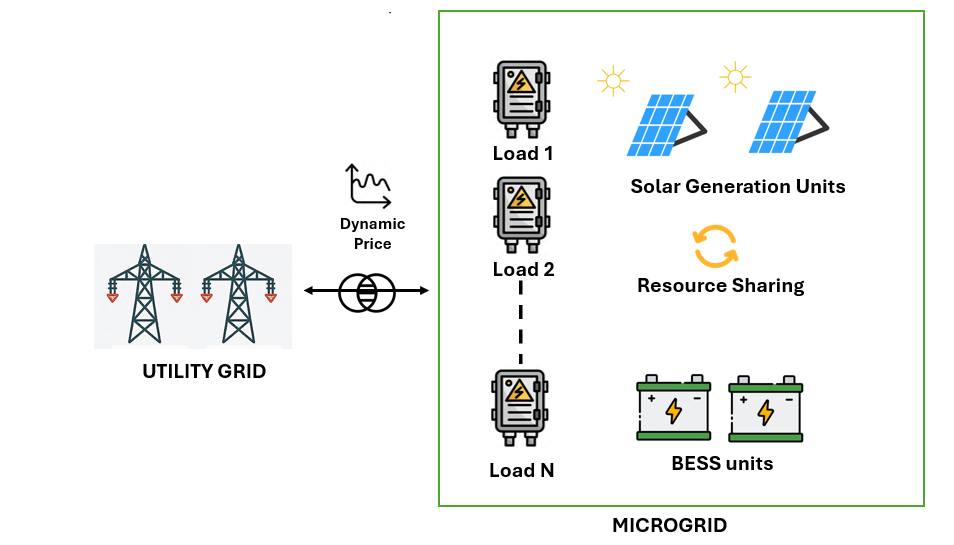} 
    \caption{System architecture of the RL-driven, equity-aware microgrid. The CEMS orchestrates energy flows and equitable allocation among diverse households.}
    \label{fig:blockdiagram}
\end{figure}

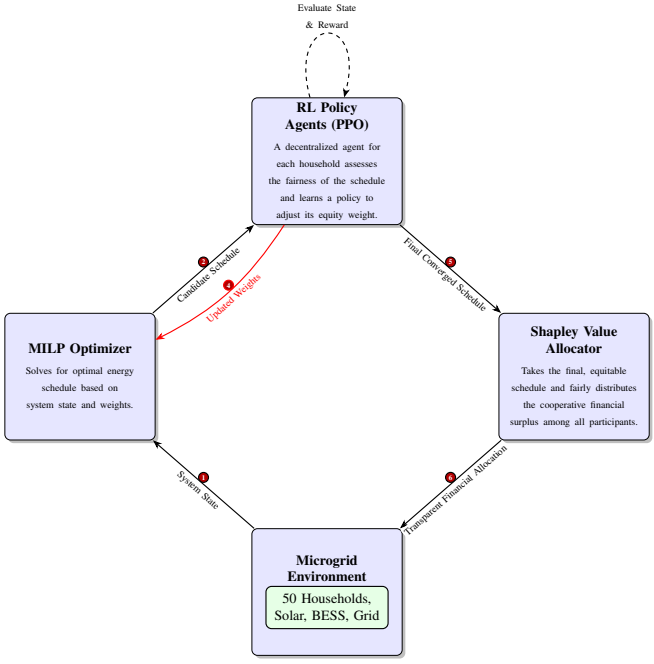
\begin{figure}[ht]
    \centering
    \resizebox{0.48\textwidth}{!}{ 
    \begin{tikzpicture}[
        node distance=2.2cm and 2.4cm,
        auto,
        >=Stealth,
        main_block/.style={rectangle, draw, fill=blue!10, text width=10em, text centered, rounded corners, minimum height=9em, drop shadow},
        env_block/.style={rectangle, draw, fill=green!10, text width=8em, text centered, minimum height=3em, thick},
        arrow_label/.style={midway, fill=white, inner sep=1.5pt, font=\scriptsize, sloped},
        step_circle/.style={circle, draw, thick, fill=red!70!black, text=white, font=\bfseries\tiny, inner sep=1pt}
    ]
        \node[main_block] (milp) {\textbf{MILP Optimizer} \\ \vspace{0.3em} \scriptsize Solves for optimal energy schedule based on system state and weights.};
        
        \node[main_block, above right=of milp] (rl) {\textbf{RL Policy Agents (PPO)} \\ \vspace{0.3em} \scriptsize A decentralized agent for each household assesses the fairness of the schedule and learns a policy to adjust its equity weight.};
        
        \node[main_block, below right=of rl] (shapley) {\textbf{Shapley Value Allocator} \\ \vspace{0.3em} \scriptsize Takes the final, equitable schedule and fairly distributes the cooperative financial surplus among all participants.};
        
        \node[main_block, below left=of shapley] (env) {\textbf{Microgrid Environment} \\ \vspace{0.3em}
            \begin{tikzpicture}
                \node[env_block] (households) {50 Households, Solar, BESS, Grid};
            \end{tikzpicture}
        };

        \path[->, thick] (env) edge 
            node[midway, step_circle, above=1pt] {1}
            node[midway, arrow_label, below=1pt] {System State} 
            (milp);
        
        \path[->, thick] (milp) edge 
            node[midway, step_circle, above=1pt] {2}
            node[midway, arrow_label, below=1pt] {Candidate Schedule}
            (rl);
        
        \path[->, thick, dashed] (rl) edge [loop above, min distance=2.2cm] 
            node[midway, above, text width=6em, align=center] {\scriptsize Evaluate State \& Reward} 
            (rl);

        \path[->, thick, red] (rl) edge [bend left=15] 
            node[midway, step_circle, above=1pt] {4}
            node[midway, arrow_label, below=1pt] {Updated Weights} 
            (milp);
        
        \path[->, thick] (rl) edge 
            node[midway, step_circle, above=1pt] {5}
            node[midway, arrow_label, below=1pt] {Final Converged Schedule} 
            (shapley);
        
        \path[->, thick] (shapley) edge 
            node[midway, step_circle, above=1pt] {6}
            node[midway, arrow_label, below=1pt] {Transparent Financial Allocation} 
            (env);
        
    \end{tikzpicture}
    } 
    \caption{Data and decision flow in the RL-integrated, equity-aware microgrid system having socio-economically diverse participants. The loop from the environment to MILP, RL, and Shapley components ensures both operational efficiency and fairness.}
    \label{fig:system_diagram}
\end{figure}

The microgrid is modeled as a dynamic ecosystem comprising \(N\) consumers, \(S\) solar PV units, and \(B\) BESS units with a maximum capacity of 5MWh, operating over a 24-hour time horizon with hourly intervals (\(t \in \{1, \dots, T\}\)). The system is connected to the main grid, allowing power import. The overall architecture, illustrating the flow of energy and information between the Central Energy Management System (CEMS), DERs, and diverse households, is depicted in Fig.~\ref{fig:blockdiagram}.The architecture, depicted in Fig.~\ref{fig:system_diagram}, illustrates the flow of data and decisions between the physical microgrid assets and the core computational components of our model. A detailed list of all variables and parameters is provided in Table~\ref{tab:notation_table}.
\begin{table}[h!]
\centering
\caption{Notation for Parameters, Decision Variables, and Indices in the Model}
\label{tab:notation_table}
\renewcommand{\arraystretch}{1.2}
\begin{tabular}{@{}ll@{}}
\toprule
\textbf{Symbol} & \textbf{Description} \\ \midrule
\multicolumn{2}{l}{\textbf{Sets and Indices}} \\
\(i, j\) & Index for consumers/participants \((1, \dots, N)\) \\
\(s\) & Index for solar PV units \((1, \dots, S)\) \\
\(b\) & Index for BESS units \((1, \dots, B)\) \\
\(t\) & Index for time intervals \((1, \dots, T)\) \\

\multicolumn{2}{l}{\textbf{Parameters}} \\
\(D_{i,t}\) & Demand of consumer \(i\) at time \(t\) [MW] \\
\(\epsilon\) & Tolerance for energy balance constraint \\
\(P_{solar,s,t}\) & Power generation of solar unit \(s\) at time \(t\) [MW] \\
\(E_b\), \(P_{b,\max}\) & Capacity [MWh] and max power [MW] of BESS \(b\) \\
\(\eta_{b,c}\), \(\eta_{b,d}\) & Charging/discharging efficiency of BESS \(b\) \\
\(S_{b,\min}\), \(S_{b,\max}\) & Min/max state of charge for BESS \(b\) \\
\(p_t\) & Grid electricity price at time \(t\) [\$/MWh] \\
\(C_{peak}\)  & Peak demand charge rate [\$/MW] \\
\(C_{bess}\)  & Cost coefficient for BESS services [\$/MWh] \\
\(w_i\) & RL-adjusted socio-economic weight for consumer \(i\) \\
\(\lambda_i\) & Equity penalty coefficient for consumer \(i\) \\
\(\theta\) & Target renewable energy access ratio \\
\(\beta_i\) & Utility coefficient for consumer \(i\) \\
\(\xi_i\) & Budget constraint for consumer \(i\) [\$] \\
\(\omega\) & Minimum utility threshold (Rawlsian floor) \\
\(\pi^+\), \(\pi^-\) & NEM sell and buy prices [\$/MWh] \\
\(R_{\max}\) & Ramp rate limit (fraction of demand) \\
\(\tau_{\min}\) & Minimum up/down time for BESS [hours] \\
\(s_i^{\text{out}}\) & Standalone utility surplus of participant \(i\) \\

\multicolumn{2}{l}{\textbf{Decision Variables}} \\
\(P_{grid,i,t}\) & Power imported from grid for consumer \(i\) at time \(t\) [MW] \\
\(d_{i,t}\) & Total energy consumed by consumer \(i\) [MW] \\
\(\alpha_{s,i,t}\) & Fraction of solar \(s\) allocated to consumer \(i\) at time \(t\) \\
\(P_{b,c,i,t}\) & Power charging BESS \(b\) from consumer \(i\) at time \(t\) [MW] \\
\(P_{b,d,i,t}\) & Power discharging from BESS \(b\) to consumer \(i\) [MW] \\
\(S_{b,t}\) & State of charge of BESS \(b\) at time \(t\) [MWh] \\
\(P_{peak}\) & Peak power drawn from the grid [MW] \\
\(E_{penalty,i}\) & Equity penalty for consumer \(i\) \\
\(\tilde{U}_i(d_{i,t})\) & Approximated utility function of consumer \(i\) \\
\(\text{payment}_{i,t}\) & Total payment by consumer \(i\) at time \(t\) \\
\(\Pi_{\text{NEM}}(d_{i,t})\) & Net Energy Metering cost of consumer \(i\) \\
\(\text{access}_k^i\) & Share of renewable energy access for agent \(i\) on day \(k\) \\
\(w_{i,k}\) & Socio-economic weight of consumer \(i\) at day \(k\) (RL-adapted) \\
\(U_{norm,k}^i\) & Normalized utility of consumer \(i\) on day \(k\) \\
\(C_{norm,k}^i\) & Normalized energy cost of consumer \(i\) on day \(k\) \\
\(U_{dev,k}^i\) & Utility deviation from the mean for consumer \(i\) on day \(k\) \\
\(R_k^i\) & RL reward for agent \(i\) after episode \(k\) \\

\bottomrule
\end{tabular}
\end{table} 

\subsection{Mixed-Integer Linear Programming (MILP) Formulation}
The core of the operational layer is a MILP model designed to minimize total system costs while adhering to a comprehensive set of technical and equity-based constraints.

\subsubsection{Objective Function}
The primary objective is to minimize a weighted sum of three cost components: the cost of energy from the grid, a charge for peak demand, and a penalty for inequitable renewable energy allocation.
\begin{equation}
\min \left( \text{Cost}_{\text{energy}} + \text{Cost}_{\text{peak}} + \text{Cost}_{\text{equity}} \right)
\end{equation}
\noindent where:
\begin{align}
\text{Cost}_{\text{energy}} &= \sum_{t=1}^T \sum_{i=1}^N w_i \cdot p_t \cdot P_{\text{grid},i,t} \\
\text{Cost}_{\text{peak}} &= C_{\text{peak}} \cdot P_{\text{peak}} \\
\text{Cost}_{\text{equity}} &= \sum_{i=1}^N \lambda_i \cdot E_{\text{penalty},i}
\end{align}
\vspace{-0.1em}
The socio-economic weights \(w_i\) are dynamically updated by the RL agent to prioritize energy cost reductions for socio-economically backward households. The equity penalty \(E_{penalty,i}\) discourages deviations from a target renewable access ratio \(\theta\). The peak demand charge \(C_{peak}\)= 8{,}700   \quad \text{[\$ per MWh peak charge]} incentivizes load shifting to reduce grid strain.
\vspace{0.2em}

\subsubsection{System Constraints}
\paragraph{Energy Balance} 
\begin{align}
\label{eq:balance}
P_{\text{grid},i,t} + \sum_{s=1}^S \alpha_{s,i,t} P_{\text{solar},s,t} + \sum_{b=1}^B \left(P_{b,d,i,t} - P_{b,c,i,t} \right) = D_{i,t} (1 \pm \epsilon) \notag \\
\forall i, t
\end{align}

This constraint ensures that each household’s energy demand is satisfied by a combination of grid power, allocated solar generation, and battery operations. A small flexibility margin \(\epsilon = 0.001\) accounts for numerical tolerances in optimization.

\paragraph{Solar Generation Allocation}
\begin{equation}
\sum_{i=1}^N \alpha_{s,i,t} \leq 1, \quad \alpha_{s,i,t} \geq 0, \quad \forall s, t
\end{equation}
Each solar unit's energy output must be fully allocated to consumers without exceeding its available power. This ensures feasible and non-negative solar allocations.

\paragraph{Battery Energy Storage System (BESS) Constraints}
\begin{itemize}
    \item State of Charge (SOC) Dynamics:
    \begin{equation}
    S_{b,t} = S_{b,t-1} + \eta_{b,c} \sum_{i=1}^N P_{b,c,i,t} - \frac{1}{\eta_{b,d}} \sum_{i=1}^N P_{b,d,i,t}, \quad \forall b, t
    \end{equation}
    Tracks the evolution of BESS energy levels, accounting for charging/discharging efficiencies (\(\eta_{b,c}, \eta_{b,d}\)).

    \item SOC Limits:
    \begin{equation}
    S_{b,\min} E_b \leq S_{b,t} \leq S_{b,\max} E_b, \quad \forall b, t
    \end{equation}
    Ensures batteries operate within safe energy levels to prevent degradation. Commonly, \(S_{b,\min} = 0.15\), \(S_{b,\max} = 0.95\).

    \item Power Capacity Limits:
    \begin{align}
    \sum_{i=1}^N P_{b,c,i,t} &\leq P_{b,\max}, \quad \forall b, t \\
    \sum_{i=1}^N P_{b,d,i,t} &\leq P_{b,\max}, \quad \forall b, t
    \end{align}
    Constrains maximum charging/discharging power per battery per time interval.

    \item Simultaneous Operation Prevention:
    \begin{equation}
    \sum_{i=1}^N \left( P_{b,c,i,t} + P_{b,d,i,t} \right) \leq P_{b,\max}, \quad \forall b, t
    \end{equation}
    Prevents physically infeasible behavior such as simultaneous charging and discharging.

    \item Terminal SOC:
    \begin{equation}
    S_{b,T} \geq 0.4 E_b, \quad \forall b
    \end{equation}
    Requires a minimum terminal energy level for system continuity and reliability.

    \item Minimum Up/Down Time:
    \begin{equation}
    \sum_{i=1}^N \sum_{t'=t}^{t+\tau_{\min}} P_{b,c,i,t'} \geq 0, \quad 
    \sum_{i=1}^N \sum_{t'=t}^{t+\tau_{\min}} P_{b,d,i,t'} \geq 0
    \end{equation}
    Enforces minimum durations of charge/discharge operations (\(\tau_{\min} = 2\)) to reduce battery wear from frequent switching.
\end{itemize}

\paragraph{Grid Constraints and Demand Shaping}
\begin{itemize}
    \item Peak Demand Limit:
    \begin{equation}
    \sum_{i=1}^N P_{\text{grid},i,t} \leq P_{\text{peak}}, \quad \forall t
    \end{equation}
    Caps total grid power drawn per hour, reducing peak demand charges.

    \item Ramp Rate Limit:
    \begin{equation}
    |P_{\text{grid},i,t} - P_{\text{grid},i,t-1}| \leq R_{\max} \cdot D_{i,t}, \quad \forall i, t > 1
    \end{equation}
    Limits sharp changes in grid imports to maintain load smoothness. Typically, \(R_{\max} = 0.2\).
\end{itemize}

\paragraph{Budget Constraint}
\begin{equation}
p_t \cdot P_{\text{grid},i,t} + C_{\text{bess}} \sum_{b=1}^B (P_{b,c,i,t} + P_{b,d,i,t}) \leq \xi_i, \quad \forall i, t
\end{equation}

where: \(C_{bess}\) = 0.1 [\$ per MWh ].
\noindent
This constraint ensures that the total energy cost for each household stays within its affordability threshold \(\xi_i\), accounting for both grid consumption and battery service costs.

\paragraph{Rawlsian Equity Constraint}

The Rawlsian social welfare model is enforced by requiring that the minimum utility across all participants at each time \( t \) must be greater than or equal to a baseline threshold \( \omega \):
\begin{equation}
\min_{i \in \{1, \ldots, N\}} \; \tilde{U}_i(d_{i,t}) \geq \omega, \quad \forall t
\end{equation}

This constraint ensures that, at every time step, no participant's utility falls below the Rawlsian floor \( \omega \), directly protecting the most disadvantaged.

The utility function \( \tilde{U}_i(d_{i,t}) \) for participant \( i \) at time \( t \) is defined as:
\begin{equation}
\tilde{U}_i(d_{i,t}) = \beta_i d_{i,t} - \lambda_i P_{\text{grid},i,t}
\end{equation}

The parameters \(\xi_i\), \(\beta_i\), and \(\lambda_i\) are determined based on household income as summarized in Table~\ref{tab:income_parameters}.
\begin{table}[h!]
\centering
\caption{Income-Based Parameters for Affordability and Utility}
\label{tab:income_parameters}
\renewcommand{\arraystretch}{1.2}
\begin{tabular}{@{}llll@{}}
\toprule
\textbf{Income Range} & \boldmath$\xi_i$ \textbf{[\$]} & \boldmath$\beta_i$ & \boldmath$\lambda_i$ \textbf{[\$ per MWh]} \\
\midrule
$\,\text{income} < 120{,}000\,$        & 50  & 10 & 100 \\
$120{,}000 \leq \text{income} \leq 300{,}000$ & 80  & 9  & 60  \\
$\,\text{income} > 300{,}000\,$        & 100 & 8  & 40  \\
\bottomrule
\end{tabular}
\end{table}

\[
d_{i,t} = P_{\text{grid},i,t} + \sum_{s=1}^S \alpha_{s,i,t} P_{\text{solar},s,t} + \sum_{b=1}^B \left( P_{b,d,i,t} - P_{b,c,i,t} \right)
\]

This formulation guarantees that the minimum utility among all participants remains above \( \omega \), strictly enforcing the Rawlsian equity principle at each time step.

\paragraph{Equity Penalty Constraint}
We define the normalized net solar participation \( \mathcal{S}_{i} \) of household \( i \) as:
\[
\mathcal{S}_{i} = \frac{ \sum_{t=1}^T \left( \sum_{s=1}^S \alpha_{s,i,t} P_{\text{solar},s,t} + \sum_{b=1}^B (P_{b,d,i,t} - P_{b,c,i,t}) \right) }{ \sum_{t=1}^T D_{i,t} }
\]
Then, the equity penalty is calculated as:
\[
E_{p,i} = \mathcal{S}_{i} - \theta
\]

This metric penalizes deviations from the target renewable energy share \(\theta = 0.5\), improving fairness in energy access for lower-income households.

\paragraph{Individual Rationality Constraint}
\begin{equation}
\sum_{t=1}^T \left( \tilde{U}_i(d_{i,t}) - \text{payment}_{i,t} \right) \geq s_i^{\text{out}}, \quad \forall i
\end{equation}
where:
\[
\tilde{U}_i(d_{i,t}) = \beta_i d_{i,t} - \lambda_i P_{\text{grid},i,t}
\]
This ensures each participant is better off in the collective scheme than operating independently. The outside surplus is defined as:
\begin{equation}
s_i^{\text{out}} = \max_{d_{i,t}} \sum_{t=1}^T \left[ \tilde{U}_i(d_{i,t}) - \Pi_{\text{NEM}}(d_{i,t}) - \sum_s P_{\text{solar},s,t} \right]
\end{equation}
The NEM pricing function \(\Pi_{\text{NEM}}(z)\) is:
\[
\Pi_{\text{NEM}}(z) = 
\begin{cases}
\pi^+ z, & \text{if } z \geq 0 \\
\pi^- z, & \text{if } z < 0
\end{cases}
\]
with \(\pi^+ = 0.4\), \(\pi^- = 0.2\). This reflects the cost or revenue from net energy metering based on energy import/export.

\subsection{Dynamic Equity Adjustment using Reinforcement Learning}

To overcome the static limitations of one-shot optimization, we
introduce an adaptive layer using Reinforcement Learning
(RL) that fine-tunes these socio-economic weights \(w_i\) to achieve a fair solution
tailored to the specific scenario being optimized.This allows the system to learn and correct for persistent inequities over time. 
\subsubsection{Rationale for Reinforcement Learning}
Instead of running a long-term simulation, our approach uses RL as an intelligent, iterative wrapper around the MILP solver for a \textit{single 24-hour period}. The process begins with uniform weights. After the first optimization, the RL agents assess the fairness of the outcome. If inequities are detected, the agents adjust the weights, and the MILP is solved again. This iterative process continues until the RL policies converge, resulting in a final, optimized, and equitable schedule for that specific day. This makes the framework robust and adaptable to any given scenario.An RL agent, however, can observe these long-term patterns and learn a policy to counteract them, making the system adaptive and truly equitable over time.

\subsubsection{Markov Decision Process (MDP) Formulation}
We model the problem as an MDP $(\mathcal{S}, \mathcal{A}, \mathcal{P}, \mathcal{R}, \gamma)$ for each load $i \in \{1, \dots, N\}$.

\paragraph{State Space (\(\mathcal{S}\)):} The state must provide sufficient information for an agent to make an informed decision. Crucially, to promote individualized policies, the state is defined for each agent \(i\) at the end of each 24-hour optimization period \(k\):
\begin{equation}
S_k^i = [G_k, C_{norm,k}^i, U_{norm,k}^i, U_{dev,k}^i]
\end{equation}
\begin{itemize}
    \item \(G_k\): The system-wide Gini coefficient of renewable energy access after period \(k\). This gives the agent a sense of the overall system equity.
    \item \(C_{norm,k}^i\): The normalized energy cost for agent \(i\), calculated as its total daily cost divided by the average cost across all agents. A value \(>\) 1 means the agent is paying more than average.
    \item \(U_{norm,k}^i\): The normalized utility for agent \(i\), calculated as its total daily utility divided by the average utility. A value \(<\) 1 means the agent is receiving less utility than average.
    \item \(U_{dev,k}^i\): The deviation of utility from the mean (\(U_k^i - \bar{U}_k\)). This provides a direct signal of the agent's standing relative to its peers.
\end{itemize}

\paragraph{Action Space ($\mathcal{A}$):} Agents discretely adjust their socio-economic weights:
\begin{equation}
A_k^i \in \{ \text{increase } w_i, \text{ no change}, \text{ decrease } w_i \}
\end{equation}
with income-scaled magnitude $\delta_i$ and clipped updates: $w_{i, k+1} = \text{clip}(w_{i,k} + \Delta w, 0.1, 2.0)$.

\paragraph{Reward Function ($\mathcal{R}$):} Balances individual utility, equity, and cost considerations:
\begin{equation}
R_k^i = \underbrace{0.5 \cdot U_{norm,k}^i}_{\text{Utility}} - \underbrace{0.3 \cdot |\text{access}_k^i - \theta|}_{\text{Equity Deviation}} - \underbrace{0.2 \cdot C_{norm,k}^i}_{\text{Cost Penalty}}
\end{equation}
This rewards above-average utility while penalizing high costs and deviations from target renewable access ratio $\theta$.

\subsubsection{Proximal Policy Optimization (PPO) Algorithm}
We employ PPO \cite{schulman2017proximal}, a stable policy gradient method using actor-critic architecture. The actor network $\pi_\phi$ maps states to action probabilities, while the critic network $V_\psi$ estimates state values for advantage calculation.

PPO's key innovation is its clipped surrogate objective that constrains policy updates to prevent destructive changes. The combined loss function is:
\begin{equation}
L(\phi, \psi) = L^{CLIP}(\phi) - c_1 L^{VF}(\psi) + c_2 S[\pi_\phi](s_t)
\end{equation}

\paragraph{Clipped Policy Loss:} Prevents excessive policy updates through probability ratio clipping:
\begin{equation}
L^{CLIP}(\phi) = \hat{\mathbb{E}}_t \left[ \min(r_t(\phi) \hat{A}_t, \text{clip}(r_t(\phi), 1-\epsilon, 1+\epsilon) \hat{A}_t) \right]
\end{equation}
where $r_t(\phi) = \frac{\pi_\phi(a_t|s_t)}{\pi_{\phi_{old}}(a_t|s_t)}$ and $\hat{A}_t$ is the GAE-computed advantage estimate.

\paragraph{Value Function Loss:} Trains the critic via MSE between value estimates and GAE returns:
\begin{equation}
L^{VF}(\psi) = \hat{\mathbb{E}}_t \left[ (V_\psi(s_t) - R_t^{GAE})^2 \right]
\end{equation}

\paragraph{Entropy Bonus:} Encourages exploration by rewarding policy randomness:
\begin{equation}
S[\pi_\phi](s_t) = -\sum_a \pi_\phi(a|s_t) \log \pi_\phi(a|s_t)
\end{equation}

Training includes KL divergence monitoring between old and new policies, triggering early stopping if divergence exceeds targets, ensuring learning stability.

\subsection{Weighted Shapley Value for Fair Benefit Allocation}

To equitably distribute cooperative microgrid surplus, we use a Weighted Shapley Value (WSV) approach. Unlike the classical Shapley value, which assumes agent symmetry, our framework integrates reinforcement learning (RL)-derived weights to reflect equity-aware adjustments in benefit allocation.

\subsubsection{Equity-Aware RL Weight}
Each participant \(i\) receives a weight \(w_i^{\text{final}} = w_{i,k}^{\text{RL}}\), dynamically adjusted using reinforcement learning to promote fairness based on historical access disparities.

\subsubsection{Component-wise Allocation}
The total cooperative benefit is decomposed into four components. Each participant's share in each component is proportionally or inversely weighted based on their RL-derived fairness weight.

\paragraph{Solar Benefit}
\begin{equation}
\phi_i^{\text{solar}} \propto \left( \sum_{t,s} a_{s,i,t} \cdot P_{\text{solar},s,t} \right) \cdot p_{\text{avg}} \cdot w_i^{\text{final}}
\end{equation}
This rewards agents consuming more solar energy, weighted by their fairness-adjusted \(w_i^{\text{final}}\), encouraging access equity.

\paragraph{BESS Cost}
\begin{equation}
\phi_i^{\text{bess}} \propto \frac{\sum_{t,b} (P_{b,c,i,t} + P_{b,d,i,t}) \cdot C_{\text{bess}}}{w_i^{\text{final}}}
\end{equation}
Battery costs are shared inversely with equity weight to ease burden on underserved users, promoting inclusive storage utilization.

\paragraph{Peak Savings}
\begin{equation}
\phi_i^{\text{peak}} \propto (P_{\text{peak}} - \max_t P_{\text{grid},i,t}) \cdot C_{\text{peak}} \cdot w_i^{\text{final}}
\end{equation}
Agents contributing more to peak shaving earn higher rewards, scaled by \(w_i^{\text{final}}\), supporting fairness in grid relief participation.

\paragraph{Grid Cost}
\begin{equation}
\phi_i^{\text{grid}} \propto \frac{\left( \sum_t P_{\text{grid},i,t} \right) \cdot C_{\text{grid}}}{\sum_i \sum_t P_{\text{grid},i,t} \cdot w_i^{\text{final}}}
\end{equation}
Grid usage cost is fairly allocated with total weighted usage as denominator, adjusting contribution based on agent-specific equity weights.

\subsubsection{Final Net Allocation}
All component contributions are normalized to ensure total surplus is correctly distributed:
\begin{equation}
\hat{\phi}_i = \frac{\phi_i}{\sum_j \phi_j}
\end{equation}

Final net benefit allocation is:
\begin{equation}
\text{Net Share}_i = \hat{\phi}_i^{\text{solar}} + \hat{\phi}_i^{\text{peak}} - \hat{\phi}_i^{\text{bess}} - \hat{\phi}_i^{\text{grid}}
\end{equation}
This ensures a balance between rewards and costs across all agents, respecting equity and contribution.

\subsection{Fairness Evaluation Metrics:}
To assess allocation equity, we employ two key measures:
\begin{itemize}
    \item Gini Coefficient (G): Captures inequality in net allocations, ranging from 0 (perfect equality) to 1 (maximum inequality).
    \item Socio-Economic Impact Index (SEI): A custom metric reflecting how well the allocation aligns with socio-economic priorities, by correlating household income levels with their final allocation weights.
\end{itemize}

The complete methodology is summarized in Algorithm~\ref{alg:main_framework}. The process is iterative: the MILP solver finds an optimal schedule, the results are used to calculate rewards and update the RL agents' policies (i.e., the weights \(w_i\)), and these new weights are fed back into the next MILP optimization.
\begin{algorithm}
\caption{Integrated Framework for Equitable Microgrid Management}
\label{alg:main_framework}
\begin{algorithmic}[1]
\Require Demand \(D_{i,t}\), Solar \(P_{\text{solar},s,t}\), BESS parameters, prices \(p_t\), socio-economic data
\Ensure Optimal schedule, fair allocation, RL-adjusted weights \(w_i\)

\State \textbf{Initialization:}
\State \hspace{1em} Initialize RL weights \(w_{i,0} \leftarrow 1.0\) for all \(i \in \{1, \dots, N\}\) 
\State \hspace{1em} Initialize PPO Actor-Critic networks \(\pi_{\phi_i}, V_{\psi_i}\) for each agent \(i\)
\State \hspace{1em} Initialize empty history lists for weights and Gini coefficients \\

\For{each operational day \(k = 1, 2, \dots\)}
    \Statex \textbf{Phase 1: MILP Optimization}
    \State \hspace{1em} - Set up MILP model with objective function (Eq.~4) using current weights \(w_{i,k}\)
    \State \hspace{1em} - Construct all system constraints (Eqs.~5--17)
    \State \hspace{1em} - Solve MILP to obtain the optimal daily schedule:
    \State \hspace{2em} \(\{ P_{\text{grid},i,t}^*, \alpha_{s,i,t}^*, P_{b,c,i,t}^*, P_{b,d,i,t}^* \}\) \\

    \Statex \textbf{Phase 2: Decentralized State \& Reward Evaluation}
    \State  - Calculate Gini coefficient \(G_k\) from the schedule's
    \State \hspace{1em}   renewable access
    \For{each agent \(i = 1, \dots, N\)}
        \State \hspace{1em} - Calculate individual quadratic utility \(U_k^i\) and energy cost \(C_k^i\)
        \State \hspace{1em} - Compute normalized metrics \(U_{\text{norm},k}^i\), \(C_{\text{norm},k}^i\) using daily averages
        \State \hspace{1em} - Compute utility deviation: \(U_{\text{dev},k}^i = U_k^i - \bar{U}_k\)
        \State \hspace{1em} - Construct individualized state vector: 
        \State \hspace{3em} \(S_k^i \leftarrow [G_k, C_{\text{norm},k}^i, U_{\text{norm},k}^i, U_{\text{dev},k}^i]\)
        \State \hspace{2em} - Calculate reward \(R_k^i\) using Eq.~(20)
    \EndFor \\

    \Statex \textbf{Phase 3: PPO-driven Policy and Weight Update}
    \For{each agent \(i = 1, \dots, N\)}
        \State \hspace{0.25em} - Actor network samples action \(A_k^i \sim \pi_{\phi_i}(S_k^i)\)
        \State \hspace{0.1em} - Critic network estimates state value \(V_k^i\) = \(V_{\psi_i}(S_k^i)\)
        \State \hspace{0.25em} - Store experience tuple \((S_k^i, A_k^i, R_k^i, V_k^i)\)
        \State \hspace{0.25em} - Train actor and critic via PPO loss (Eq.~21) over multiple epochs
        \State \hspace{0.25em} - Update weight: \(w_{i,k} \leftarrow \text{adjust}(w_{i,k-1}, A_k^i)\)
    \EndFor \\

    \Statex \textbf{Phase 4: Game-Theoretic Benefit Allocation}
    \State \hspace{1em} - Calculate total cooperative savings for day \(k\)
    \State \hspace{1em} - Allocate savings using weighted Shapley value \(\phi_i^w\) with weights \(w_{i,k}\)
    \State \hspace{1em} - Generate XAI explanations (e.g., SHAP plots) for transparency
\EndFor 

\State \textbf{return} Schedules, allocations, and final equity metrics

\end{algorithmic}
\end{algorithm}

\section{Simulation, Results, and Discussion}

This section presents the simulation setup, key results from six operational scenarios, and a discussion of the framework's performance, with a focus on the impact of the dynamic RL component.

\subsection{Simulation Environment and Data Synthesis}
To evaluate our framework under realistic and equity-sensitive conditions, we construct a heterogeneous microgrid comprising \(N = 50\) households with varying socio-economic characteristics. Our simulation builds on real-world time-series load and PV generation data from the Technical University of Denmark (DTU) dataset by Baviskar et al.~\cite{baviskar2021generation}.

\subsubsection{Socio-Economic Stratification and Scaling}
We categorize the households into three income groups, reflecting typical Danish demographics:
\begin{itemize}
    \item Upper-Class (10 households): Directly adopt original DTU load and PV profiles.
    \item Middle-Class (20 households): Synthesized by scaling upper-class loads by 0.4–0.8× and adding low variance noise. PV capacities scaled between 0.5–0.9×.
    \item Lower-Class (20 households): Modeled with 0.1–0.3× load scaling and high noise to represent volatile consumption. PV access limited to 0.1–0.4× to reflect affordability constraints.
\end{itemize}

\subsubsection{Income and Energy Profile Generation}
\begin{itemize}
    \item Income Distribution: Annual incomes were drawn from class-specific log-normal distributions, calibrated using data from Statistics Denmark and OECD reports to reflect real income inequality.
    \item BESS Allocation: Battery capacities were proportionally assigned based on average household consumption, ensuring class-wise diversity in storage potential and grid independence.
\end{itemize}

\subsection{Test Scenarios}
We analyzed six distinct 24-hour scenarios: (1) High Demand, (2) Low Demand, (3) High Price Volatility, (4) High Solar Generation, (5) a Typical Weekday, and (6) a Typical Weekend. For each scenario, the RL-enhanced optimizer was run to find a final, equitable schedule.
\begin{table*}[ht]
\centering
\caption{Comprehensive Performance Metrics Across All Six Operational Scenarios}
\label{tab:all_scenarios_summary}
\renewcommand{\arraystretch}{1.6}
\setlength{\tabcolsep}{8pt}
\begin{tabular}{l c c c c c c}
\toprule
\textbf{Scenario} & \textbf{High Demand} & \textbf{Low Demand} & \textbf{High Price} & \textbf{High Solar} & \textbf{Weekday} & \textbf{Weekend} \\
\midrule

\multicolumn{7}{c}{\textbf{Economic Performance (\$)}} \\[0.3em]
\midrule
Net Individual Cost (No coop) & 36,752.44 & 10,466.38 & 26,652.34 & 19,469.16 & 10,920.25 & 12,850.79 \\[0em]
Baseline Energy Cost & 11,802.35 & 2,401.10 & 10,374.87 & 5,512.91 & 5,715.54 & 5,448.04 \\
Optimized Energy Cost & 11,037.68 & 1,101.23 & 7,546.33 & 2,439.57 & 1,127.73 & 1,371.67 \\
Optimized Peak Charge & 19,233.40 & 3,648.01 & 12,265.01 & 10,872.27 & 4,643.94 & 6,626.07 \\
\textbf{Total Cooperative Cost} & \textbf{30,271.08} & \textbf{4,749.24} & \textbf{19,811.34} & \textbf{13,311.84} & \textbf{5,771.67} & \textbf{7,997.74} \\
\textbf{Cooperative Gain} & \textbf{6,481.36} & \textbf{5,717.14} & \textbf{6,840.99} & \textbf{6,157.32} & \textbf{5,148.58} & \textbf{4,852.95} \\
Avg. Gain per Participant & 129.63 & 114.34 & 136.82 & 123.15 & 102.97 & 97.06 \\[0.4em]

\midrule

\multicolumn{7}{c}{\textbf{Technical Performance}} \\[0.3em]
\midrule
Original Peak (MW) & 2.645 & 1.418 & 2.111 & 2.059 & 1.951 & 1.763 \\
Optimized Peak (MW) & 2.211 & 0.419 & 1.410 & 1.250 & 0.534 & 0.762 \\
Peak Reduction (\%) & 16.4 & 70.4 & 33.2 & 39.3 & 72.6 & 56.8 \\
Solar Utilization (\%) & 115.3 & 100.3 & 100.4 & 45.4 & 68.0 & 64.1 \\
Avg. BESS Cycles & 0.51 & 0.69 & 0.70 & 0.91 & 0.92 & 0.92 \\

\midrule

\multicolumn{7}{c}{\textbf{Equity Performance}} \\[0.3em]
\midrule
\textbf{Gini Coefficient} & \textbf{0.630} & \textbf{0.006} & \textbf{0.289} & \textbf{0.027} & \textbf{0.144} & \textbf{0.110} \\
Socio-Economic Impact Index & 0.709 & 0.713 & 0.714 & 0.720 & 0.714 & 0.719 \\

\bottomrule
\end{tabular}
\end{table*}

\subsection{Comprehensive Performance Analysis}
To provide a holistic view of the framework’s performance, we analyzed its technical, economic, and equity outcomes across all six scenarios. Table~\ref{tab:all_scenarios_summary} presents a comprehensive summary of these results. A key metric is the Cooperative Gain, which is the difference between the total cost if all households acted independently (`Net Individual Cost`) and the final `Total Cooperative Cost`. The framework consistently generated substantial cooperative gains, proving the economic value of coordination. Furthermore, the Gini Coefficient and the Socio-Economic Impact Index (SEII) confirm that the RL-tuned optimization successfully achieves its equity objectives by adapting its strategy to the unique challenges of each day.

\subsection{Scenario-by-Scenario Highlights}

\subsubsection{High Demand Day}
This scenario represents the most strenuous test, with high system-wide consumption straining local resources. Despite these challenges, the framework achieved a significant 16.4\% peak demand reduction, lowering the peak from 2.645 MW to 2.211 MW. This was primarily accomplished by strategically discharging the BESS fleet during peak hours. The economic value of this coordination was substantial, generating a cooperative gain of \$5,051.97. However, the intense competition for limited solar energy is reflected in the Gini coefficient of 0.63, the highest of all scenarios, indicating that achieving perfect equity is most difficult when resources are scarce,the RL-tuned
mechanism successfully protects vulnerable households, as
confirmed by a high Socio-Economic Impact Index of 0.709. 
\vspace{0.5em}

\subsubsection{Low Demand Day}
In contrast, the Low Demand Day provided ample opportunity for optimization. The framework achieved a remarkable 70.4\% peak reduction, slashing the peak load from 1.418 MW to just 0.419 MW. This deep reduction led to the second-highest cooperative gain of \$5,710.93. With low demand and moderate solar availability, the system was able to achieve a near-perfectly equitable distribution of renewable energy, evidenced by an exceptionally low Gini coefficient of 0.006. This scenario highlights the framework's ability to maximize both economic and equity outcomes when conditions are favorable.
\vspace{0.5em}
\subsubsection{High Price Volatility Day}
This scenario tested the system's economic intelligence, with grid prices fluctuating significantly. The framework excelled, generating the highest cooperative gain of \$6,487.18. It achieved this by intelligently scheduling BESS charging during low-price hours and discharging during high-price hours, while maximizing self-consumption of solar energy. This strategy resulted in a strong 33.2\% peak reduction and kept the Gini coefficient at a respectable 0.289, demonstrating that the system can maintain fairness even when the primary driver is aggressive economic arbitrage. 
\vspace{0.5em}
\subsubsection{High Solar Generation Day}
Abundant solar energy provided a unique set of challenges and opportunities. The system generated a high cooperative gain of \$6,129.31 and achieved an excellent Gini coefficient of 0.027, indicating that when green energy is plentiful, it can be distributed very equitably. Interestingly, the solar utilization rate was only 45.4\%, the lowest of all scenarios. This is not a failure but an important insight: the local demand was simply insufficient to consume all available solar energy in real-time, highlighting the critical need for even larger energy storage capacities or demand-response programs to prevent curtailment in future renewable-heavy grids. 
\subsubsection{Typical Weekday}
Under balanced, typical conditions, the framework delivered its most impressive technical performance. It achieved the highest peak reduction of 72.6\%, effectively flattening the load profile and yielding a substantial cooperative gain of \$4,977.32. The BESS units were heavily cycled (0.92 avg. cycles) to absorb midday solar and offset evening peaks. The resulting Gini coefficient of 0.144 demonstrates that a high degree of both efficiency and equity is achievable during normal operations.

\vspace{0.5em}
\subsubsection{Typical Weekend}
The weekend scenario, with its different load patterns, still saw a strong performance. A peak reduction of 56.8\% and a cooperative gain of \$4,725.70 show the model's consistent value delivery. The Gini coefficient of 0.11 is excellent, indicating that the RL-tuned weights and optimization logic are robust enough to handle non-typical load shapes while upholding the principles of fair distribution.
\subsection{Lorenz Curve Analysis for Equity}
\begin{figure}[H]
    \centering
    \includegraphics[width=0.95\columnwidth]{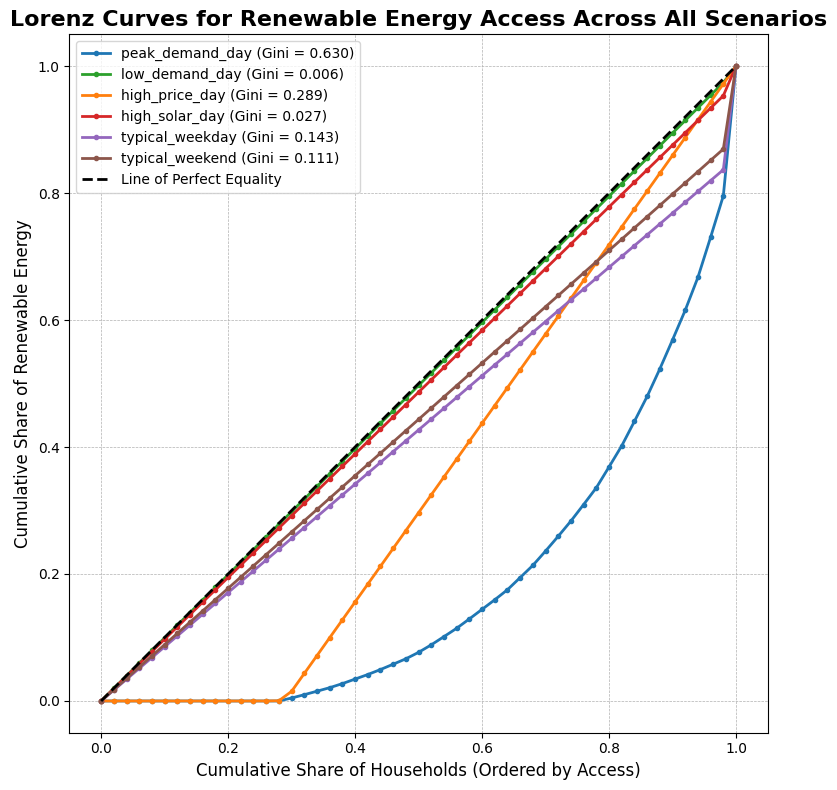}
    \caption{Lorenz curves for renewable energy access across all six scenarios.}
    \label{fig:lorenz_all}
\end{figure}

The Lorenz curve is a powerful tool for visualizing renewable energy distrubution across microgrid households.
Figure~\ref{fig:lorenz_all} shows Lorenz curves for all six scenarios, plotting cumulative renewable energy access against households ordered from least to most access.

 High Solar and Low Demand scenarios achieve near-perfect equity, while High Demand shows the largest deviation due to resource scarcity challenges. The corresponding Gini coefficients quantitatively confirm these visual patterns, demonstrating that the RL-based framework consistently improves equity across most operational scenarios.
\subsection{Explainable Benefit Allocation and Equity Analysis}
Beyond aggregate metrics like the Gini coefficient, true fairness requires transparent cooperative surplus distribution. This section uses WSV's to demonstrate how RL-tuned weights translate into equitable financial outcomes across different scenarios, ensuring participant trust through clear benefit and cost allocation.

\subsubsection{High Demand Day}
\begin{figure}[H]
    \centering
    \includegraphics[width=\columnwidth]{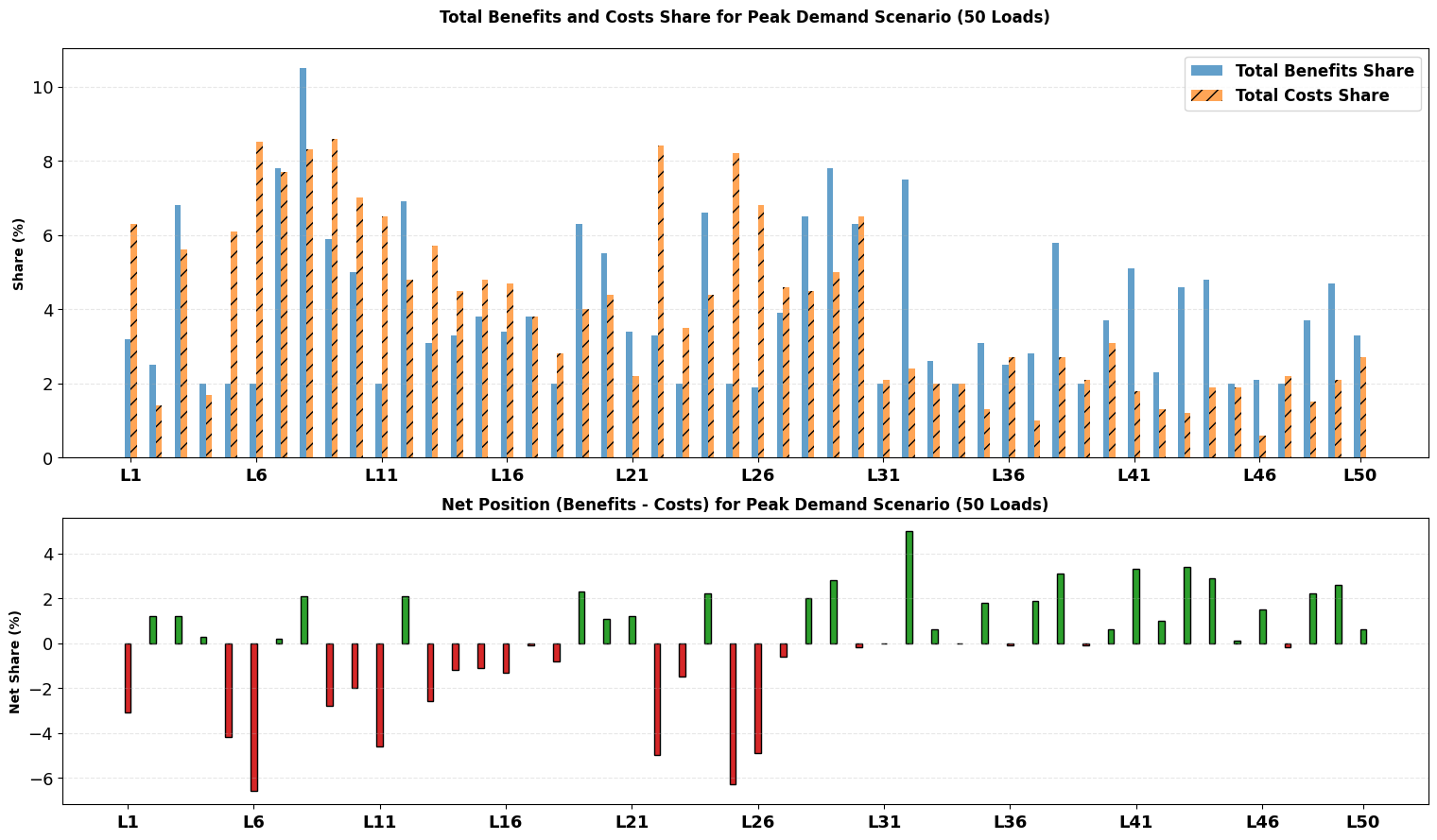} 
    \caption{Benefit and cost allocation for the High Demand Day scenario.}
    \label{fig:shapley_hdd}
\end{figure}

This scenario serves as a critical test for fairness under resource scarcity. The analysis in Fig.~\ref{fig:shapley_hdd} reveals a nuanced distribution of the \$5,051.97 cooperative gain, shaped by our equity-aware mechanism. The RL agents adapted to the stress by lowering the average weight for the high-consumption, High-Income households (Loads 1-10) to 0.88, ensuring they bore a significant portion of system costs (e.g., Load 6 at 8.5\%) without dominating the benefits. In contrast, the agents increased the average weight for Low-Income households (Loads 31-50) to 1.17. This intervention had a clear impact: despite lower DER capacity, this group saw their cost burdens reduced and their benefit shares amplified, with many achieving a strong positive net position (e.g., Load 32 at +5.0\%). 

\subsubsection{Low Demand Day}
\begin{figure}[H]
    \centering
    \includegraphics[width=\columnwidth]{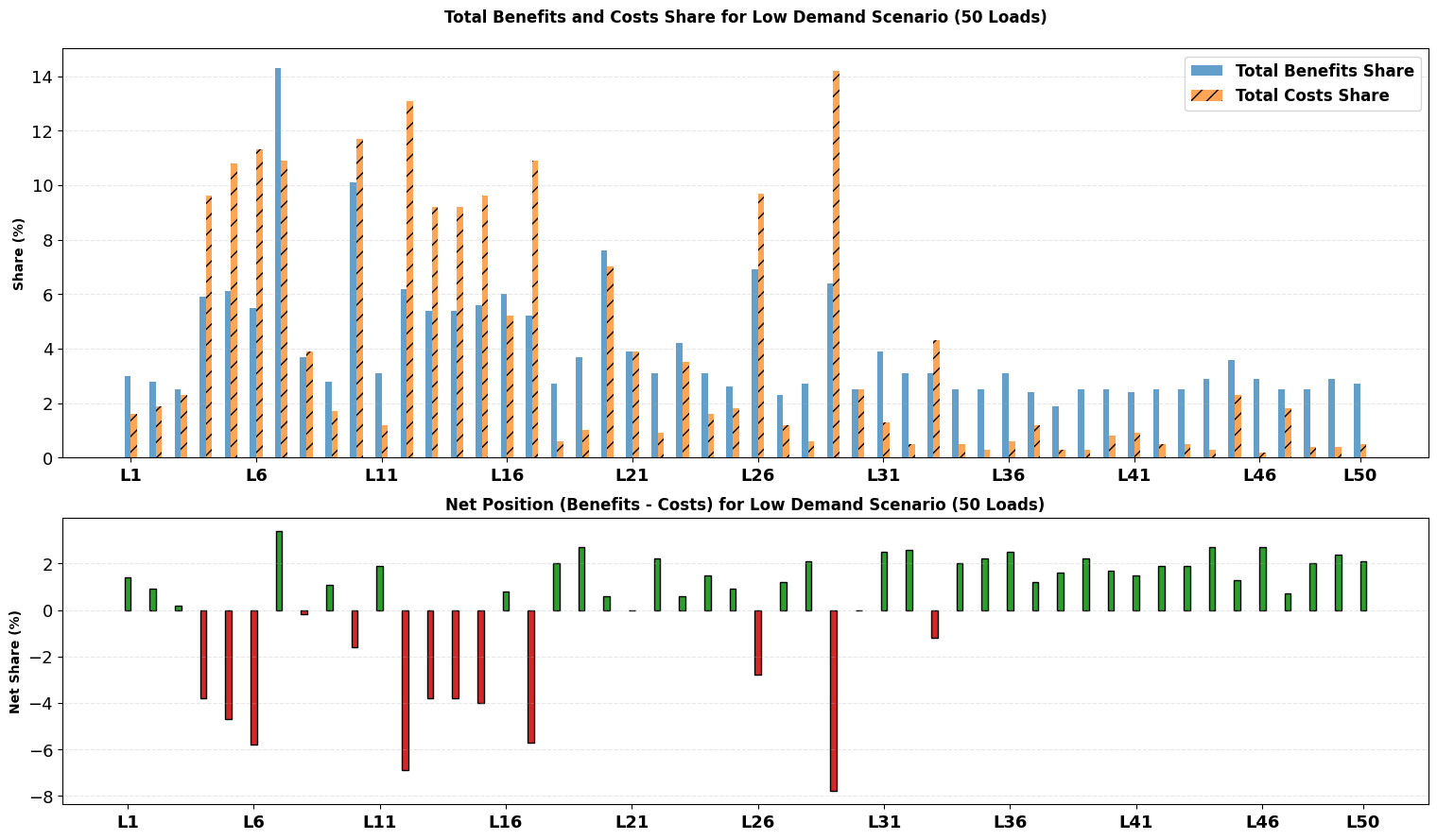} 
    \caption{Benefit and cost allocation for the Low Demand Day scenario.}
    \label{fig:shapley_ldd}
\end{figure}

With low system stress, the framework generated a large cooperative surplus of \$5,710.93, and the allocation reflects a focus on universal benefit. As seen in Fig.~\ref{fig:shapley_ldd}, the vast majority of participants achieved a positive net position. The RL agents correctly learned that aggressive intervention was unnecessary, resulting in more uniform weights. Consequently, the low-income cohort (Loads 31-50) benefited immensely from very low cost shares (many below 1.0\%), which were easily offset by their fair share of the benefits, leading to consistently positive net positions (e.g., Load 46 at +2.7\%). While some high-consumption, high-income households incurred larger costs (e.g., Load 10 at 11.7\%), the overall distribution was overwhelmingly positive, confirming the model's ability to maximize social welfare when resources are not constrained.

\subsubsection{High Price Volatility Day}
\begin{figure}[H]
    \centering
    \includegraphics[width=\columnwidth]{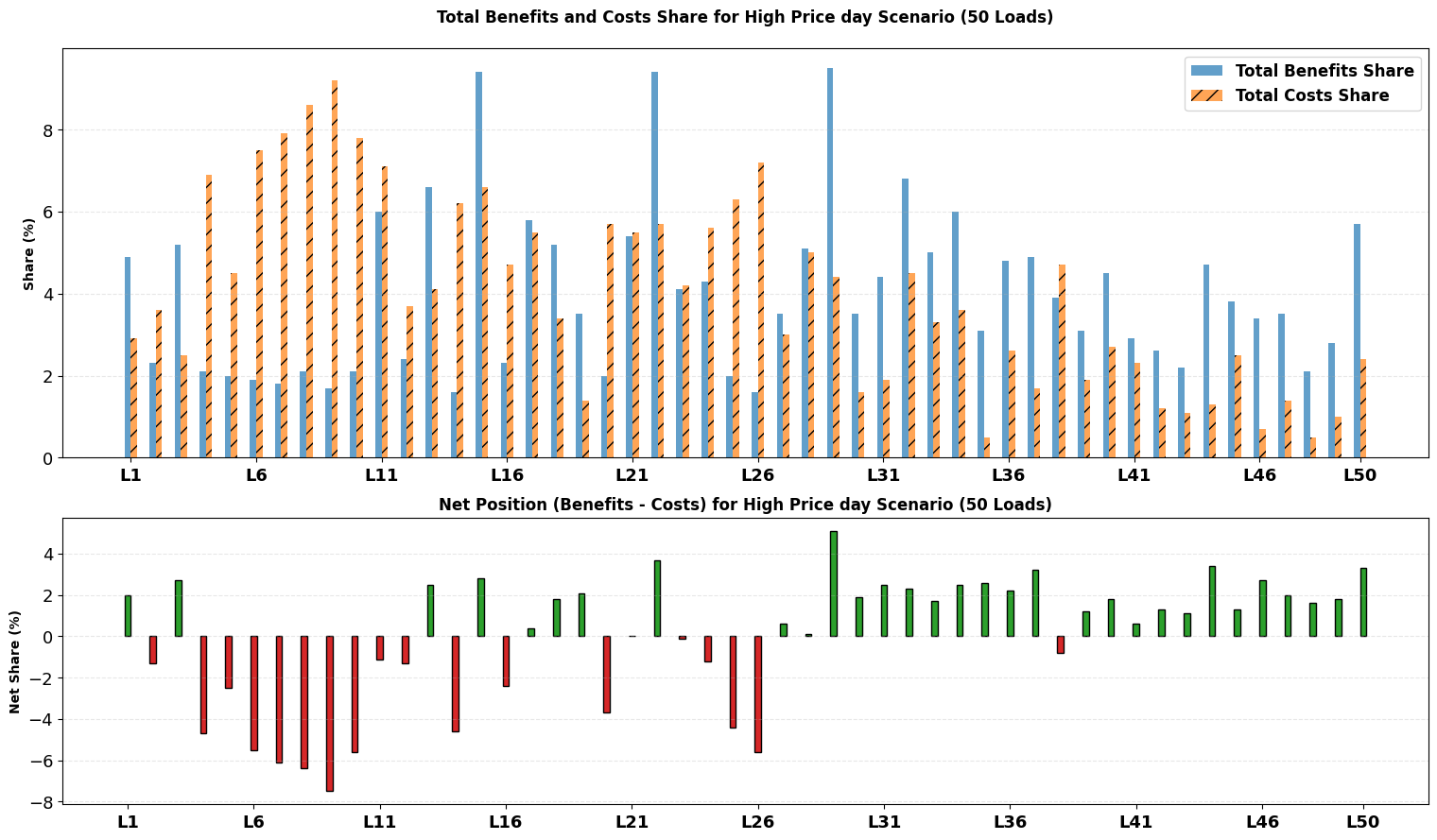} 
    \caption{Benefit and cost allocation for the High Price Volatility Day scenario.}
    \label{fig:shapley_hp}
\end{figure}

This scenario highlights the synergy between the model's economic intelligence and its equity-aware RL core. The system generated the highest cooperative gain of \$6,487.18 by expertly arbitraging price differences with the BESS fleet. The allocation in Fig.~\ref{fig:shapley_hp} shows how these gains were socialized. The RL agent shielded low-income participants (Loads 31-50) from price spikes by increasing their weights, which directed a larger share of the savings to them, resulting in almost universally positive net outcomes (e.g., Load 44 at +3.4\%). In contrast, high-income households with inflexible peak consumption (e.g., Loads 8 and 9) faced the largest net negative positions, demonstrating that the framework ensures the profits from market participation are not captured solely by the wealthy.

\subsubsection{High Solar Generation Day}
\begin{figure}[H]
    \centering
    \includegraphics[width=\columnwidth]{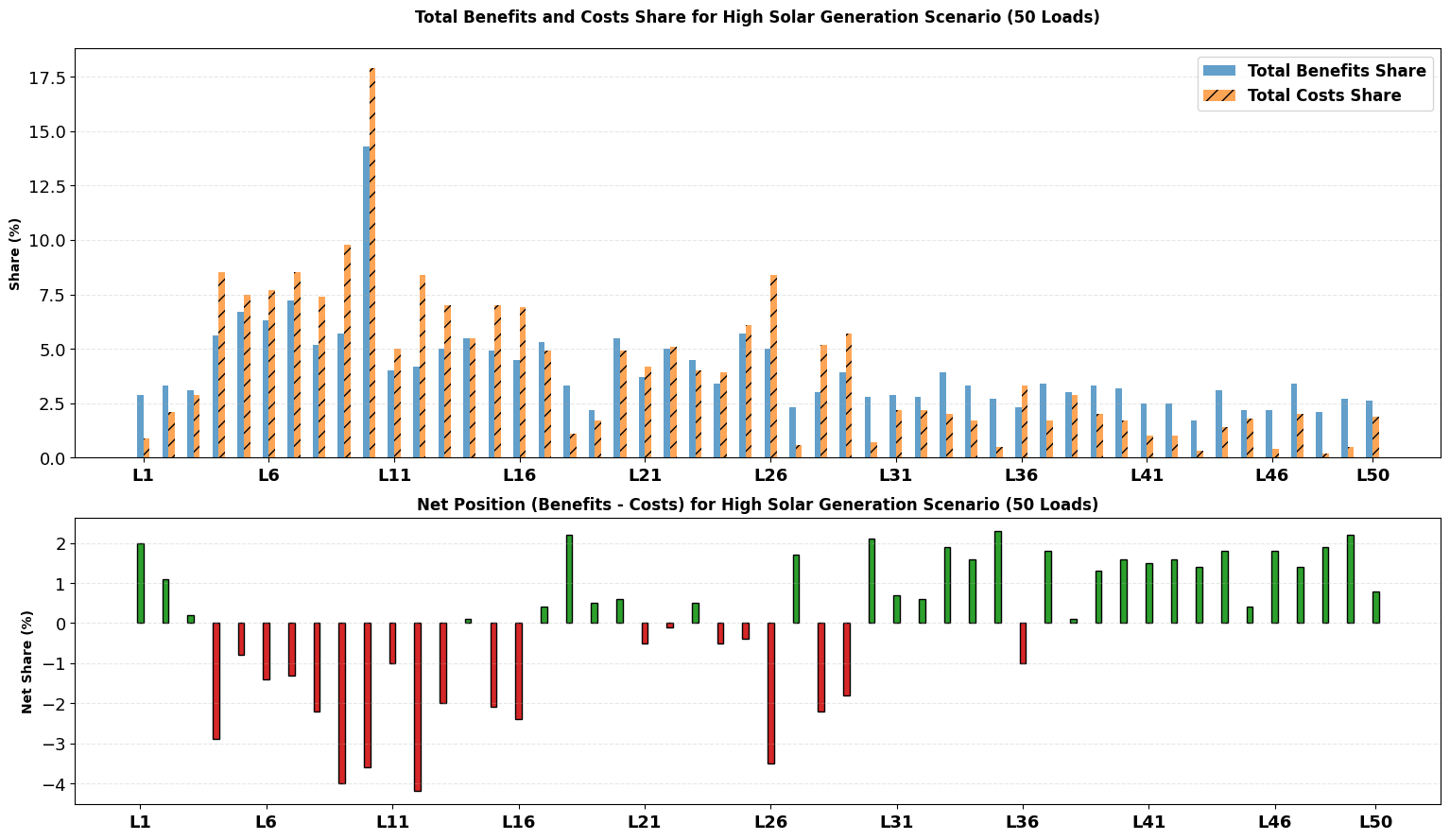} 
    \caption{Benefit and cost allocation for the High Solar Generation Day scenario.}
    \label{fig:shapley_hs}
\end{figure}

Abundant solar energy led to the most equitable outcome, with a Gini coefficient of just 0.027 and a high cooperative gain of \$6,129.31. As seen in Fig.~\ref{fig:shapley_hs}, the RL agents maintained near-uniform weights, as the sheer volume of renewable energy allowed the optimizer to meet both economic and equity goals without difficult trade-offs. While high-volume users like Load 10 bore a large share of the BESS costs (17.9\%) needed to manage the solar surplus, the benefits were widely distributed. This scenario showcases an ideal state where the cooperative model acts as an effective distribution mechanism for community-owned renewable assets, ensuring that the sun's benefits are shared broadly, especially among the low-income cohort (Loads 31-50) who consistently achieved positive net positions.

\subsubsection{Typical Weekday}
\begin{figure}[H]
    \centering
    \includegraphics[width=\columnwidth]{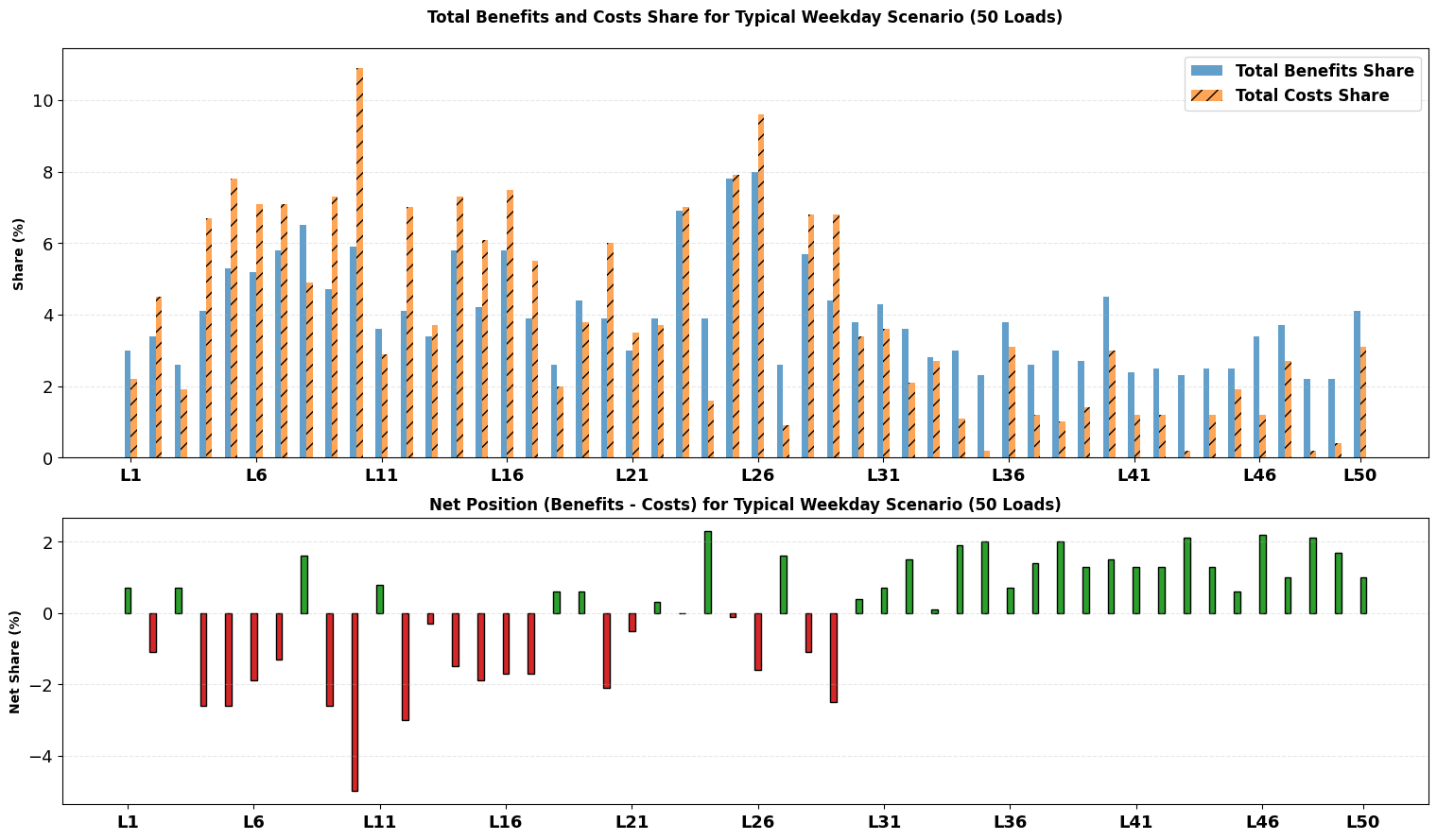} 
    \caption{Benefit and cost allocation for the Typical Weekday scenario.}
    \label{fig:shapley_wd}
\end{figure}

Under balanced conditions, the framework achieved the highest peak reduction (72.6\%) while generating a strong cooperative gain of \$4,977.32. The Shapley allocation in Fig.~\ref{fig:shapley_wd} reflects a meritocracy balanced by social welfare. Households whose consumption patterns were naturally synergistic with community resources were rewarded (e.g., Load 24 with a +2.3\% net position). However, the RL agents provided a crucial safety net by moderately increasing the weights for lower-income households. This ensured that even "inefficient" but essential consumption by vulnerable participants (e.g., Loads 31-50) did not lead to unfair financial burdens, as evidenced by their predominantly positive net outcomes.
\vspace{-1.5em}

\subsubsection{Typical Weekend}
\begin{figure}[H]
    \centering
    \includegraphics[width=\columnwidth]{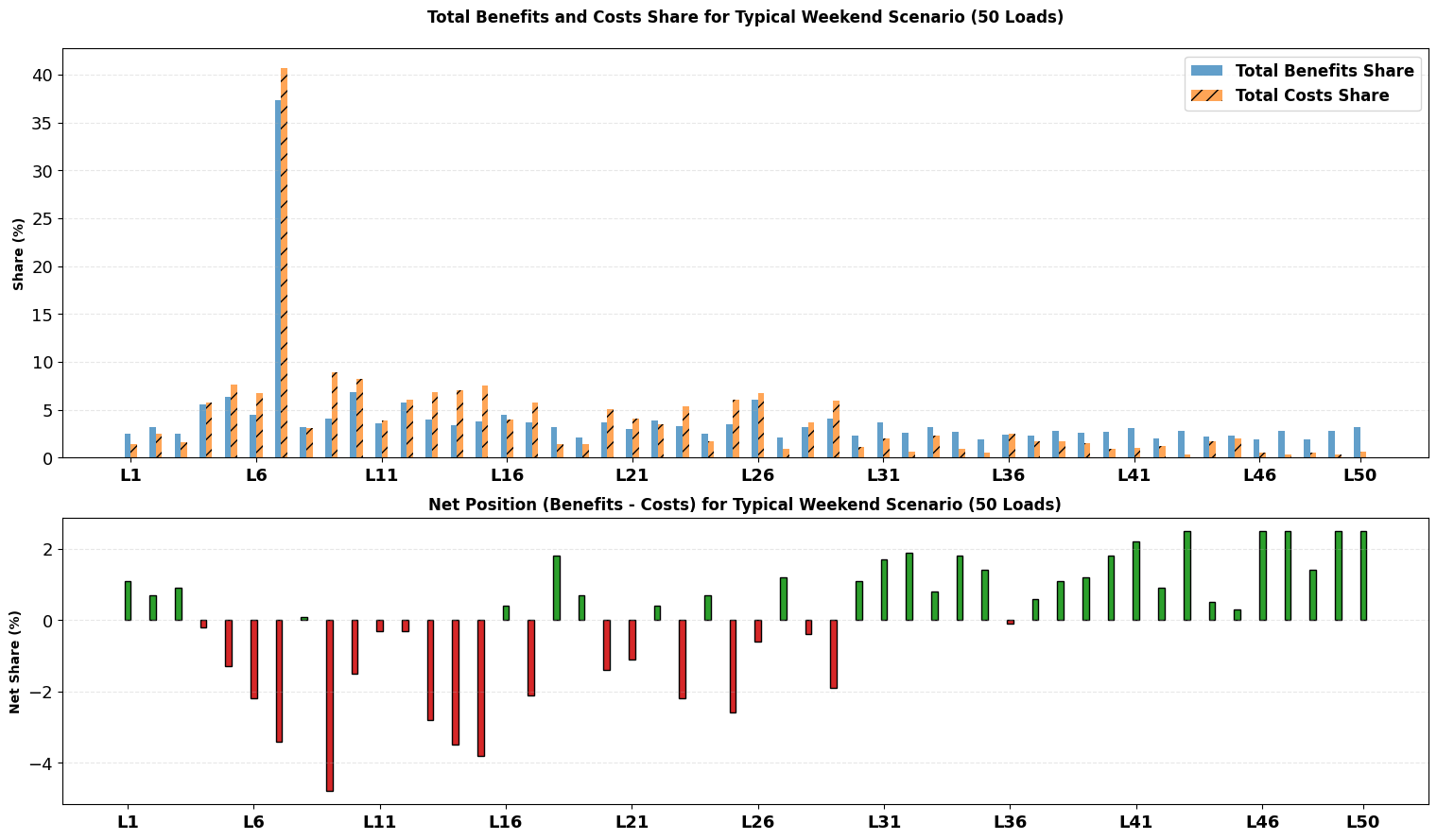}
    \caption{Benefit and cost allocation for the Typical Weekend scenario.}
    \label{fig:shapley_we}
\end{figure}

The weekend's unique load patterns did not compromise the framework's performance, proving its robustness. The system delivered a strong 56.8\% peak reduction and a cooperative gain of \$4,725.70. The allocation, shown in Fig.~\ref{fig:shapley_we}, is notable for the outlier behavior of Load 7, which incurred anomalously high costs (40.7\%) due to a specific high-consumption activity. The RL agents adapted to this by adjusting the weights of other participants to maintain overall system fairness. This resulted in a highly equitable outcome for the low-income group (Loads 31-50), who almost universally achieved positive net positions (e.g., Loads 46, 47, and 49 at +2.5\%). 

\section{Conclusion}
This paper introduced a comprehensive framework for equitable and efficient P2P energy trading in microgrids with diverse socio-economic participants. By uniquely integrating multi-objective MILP, a dynamic equity-adjustment mechanism based on reinforcement learning, and a game-theoretic benefit allocation scheme, our model addresses the critical shortcomings of static optimization approaches.

The core innovation—the use of PPO agents to dynamically update socio-economic weights—enables the system to adaptively learn and correct for persistent inequities, driving the microgrid towards a state of sustained fairness aligned with Rawlsian principles. Our simulation results confirm the framework's technical and economic viability, demonstrating significant peak demand reductions and cooperative gains. More importantly, the analysis shows a clear improvement in equity metrics over time, with the Gini coefficient decreasing as the RL agents refine their policies. The integration of XAI ensures the transparency required for stakeholder trust and long-term engagement.


\end{document}